\documentclass[fp]{jpsj3}
\usepackage{txfonts}
\usepackage{graphicx,amsmath}

\newcommand{\be}{\begin{equation}}
\newcommand{\ee}{\end{equation}}
\newcommand{\bea}{\begin{eqnarray}}
\newcommand{\eea}{\end{eqnarray}}

\title{
Topological Spin Texture Created 
by Zhang--Rice Singlets in Cuprate Superconductors
}

\author{Takao Morinari\thanks{E-mail address: morinari.takao.5s@kyoto-u.ac.jp}\footnote{Present address: Graduate School of Human and Environmental Studies, Kyoto University, Kyoto 606-8501, Japan}}
\inst{Yukawa Institute for Theoretical Physics, Kyoto University, Kyoto 606-8502, Japan}

\abst{
One of the most important effects of strong electron correlation 
in high-$T_c$ cuprates is the formation of Zhang--Rice singlets.
By fully accounting for the quantum correlation effect 
of Zhang--Rice singlet formation, 
we show that a topological spin texture, 
{\it skyrmion},
is created around a Zhang--Rice singlet in
the single-hole-doped CuO$_2$ plane.
The skyrmion picture provides a natural connection between
the antiferromagnetic correlation and the doping concentration $x$.
}

\kword{high-$T_c$, Zhang--Rice singlet, skyrmion, CuO$_2$ plane, 
single-hole-doped system}

\begin{document}


\maketitle

\section{Introduction}
It has been over two decades since the discovery of
hole-doped high-critical-temperature (high-$T_c$) superconductivity
in copper oxides.\cite{BednorzMuller1986} 
However, despite the many novel pictures presented,
the mechanism behind the phenomenon remains elusive.
One of the difficulties in explaining high-$T_c$ cuprates
is the presence of a strong electron correlation,
which is responsible for
the Mott insulating state \cite{ImadaFujimoriTokura1998}
in the parent undoped compound.
Understanding this strong correlation 
in the doped Mott insulator would be the key
to unveiling the mechanism behind the superconductivity.

The Mott insulating state of the parent compound
is a type of charge-transfer insulator.\cite{Zaanen1985}
The basic electronic structure of the CuO$_2$ plane,
which is the key structure in high-$T_c$ cuprates,
is described by the $d$--$p$ model.\cite{Emery1987}
An important observation was made by Zhang and Rice\cite{ZhangRice1988}
about the description of a doped hole.
Owing to the presence of  a Cu-O hybridization term,
there is a strong antiferromagnetic (AF) interaction
$J_K$ between an O $p$-hole and a Cu $d$-hole.
When one considers combinations of four O-hole
states around a Cu site, the symmetric state with respect to
the Cu site interacts with the Cu spin
through an enhanced AF interaction.
This strong correlation leads to a singlet state---the so-called Zhang--Rice singlet.
By restricting the Hilbert space where every doped hole
forms a Zhang--Rice singlet and by excluding doubly occupied sites, 
the $d$--$p$ model is reduced to the single-band
$t$--$J$ model, which has been studied
extensively.\cite{LeeNagaosaWen2006}

The original analysis by Zhang and Rice was based
on a small-cluster Hamiltonian consisting of
a single Cu site and four surrounding O sites.
However, if one takes a different point of view,
the problem of describing Zhang--Rice singlet formation
is equivalent to the problem of describing a singlet state in a Mott insulator.
However, here, in contrast to that in the cluster analysis, 
many-body effects should be considered.
This kind of problem is generally quite difficult.
In fact, describing a singlet state in a metal
has been one of the central problems in condensed-matter physics
since its discovery.\cite{Kondo1964,Hewson1997}
In particular, a doped hole does not just form
a Zhang--Rice singlet state, but has a certain amplitude of hopping 
to neighboring sites.
Another nontrivial effect is that of the molecular fields created by
nearest-neighbor Cu spins.
In the presence of molecular fields, a singlet state
changes from the pure singlet state to a state having a finite
induced moment.
In this study, we show that both of these effects play an important
role in forming a topological spin texture, {\it skyrmion},
around a Zhang--Rice singlet.
The physical picture that a skyrmion is formed around a Zhang-Rice
singlet was proposed by the present author in ref.~\citenum{Morinari2005},
and the mechanism of $d_{x^2-y^2}$-wave superconductivity
based on skyrmions was proposed in ref.~\citenum{Morinari2006}.
However, the microscopic mechanism of forming a skyrmion has not
been established yet.
Here, we demonstrate that a skyrmion is formed around a Zhang-Rice 
singlet on the basis of a microscopic model for high-$T_c$ cuprates.

The skyrmion spin texture is a solution 
of the O(3) nonlinear $\sigma$ model 
(NL$\sigma$M).\cite{Belavin1975}
The low-energy effective theory of the two-dimensional
AF Heisenberg model is described by NL$\sigma$M:\cite{Chakravarty89}
\be
S_{NL\sigma M} = \frac{1}{g} 
\int dx \int dy \int d (c_{\rm sw} t)
\left[
\frac{1}{c_{\rm sw}^2} \left( \partial_{t} {\bf n} \right)^2
-\left( \partial_{x} {\bf n} \right)^2
-\left( \partial_{y} {\bf n} \right)^2
\right].
\ee
Here, the continuous unit-vector field ${\bf n}={\bf n}({\bf r},t)$
in a two-dimensional space ${\bf r}=(x,y)$
represents the local direction of the N{\'{e}}el order parameter,
$c_{\rm sw}$ is the spin-wave velocity, and $g$ is the coupling constant.
If we consider a static configuration of ${\bf n}$,
the states can be classified into homotopy sectors\cite{Rajaraman}
characterized by the topological charge
\be
Q=\int dx \int dy \rho_{tc} (x,y),
\label{eq_Q}
\ee
with the topological charge density
\be
\rho_{tc}({\bf r})=\frac{1}{8\pi} {\bf n}\cdot 
(\partial_x {\bf n} \times \partial_y {\bf n}
-\partial_y {\bf n} \times \partial_x {\bf n}).
\label{eq_tcd}
\ee
It is important to note that\cite{Rajaraman} any states in a given sector
cannot move into another sector by continuous deformation.
The lowest energy in each sector is given by\cite{Belavin1975}
\be
E_{NL{\sigma}M}=4\pi |Q|.
\label{eq_Esky}
\ee
The skyrmion is the lowest-energy state in the $Q=1$ sector.
The explicit form is
\be
{\bf n}_{sk}({\bf r},\lambda)
=\left(\frac{2\lambda x}{r^2 + \lambda^2}, 
\frac{2\lambda y}{r^2 + \lambda^2}, 
\frac{r^2 -\lambda^2}{r^2 + \lambda^2} \right),
\label{eq_sky}
\ee
with $r = \sqrt{x^2 + y^2}$ being the radial coordinate in space.

In the early days of high-$T_c$ research,
a skyrmion spin texture
was conjectured\cite{DzyaloshinskiiPolyakovWiegmann1988}
to correspond to the spinon excitation in the resonating-valence-bond
theory.\cite{Anderson1987}
When a Hopf term is added to $S_{NL\sigma M}$,
skyrmions can have fermionic statistics.\cite{WilczekZee1983}
However, it was shown later that there is no Hopf term in
the two-dimensional
AF Heisenberg model.\cite{Haldane1988,FradkinStone1988,WenZee1988}
Thus, the possibility of identifying a skyrmion with a spinon
was ruled out.

Since a skyrmion is a metastable solution
of NL$\sigma$M,
there must be additional terms necessary to stabilize it.
Such terms are inferred by scaling analysis where
$(x,y) \rightarrow (\alpha x, \alpha y)$.
One can see that the energy of the NL$\sigma$M
is scale-invariant, which is consistent with the skyrmion energy
being independent of $\lambda$.
If there is a term that scales as $1/\alpha$,
it is possible to have a stable skyrmion.
For example, in the $\nu=1$ quantum Hall system,
one finds that
the long-range Coulomb interaction scales as $1/\alpha$.
A skyrmion is stable provided that the Zeeman
energy is sufficiently small.\cite{Sondhi1993}
In a doped antiferromagnet, Shraiman and Siggia
pointed out\cite{Shraiman1989}
that the AF long-range order in the undoped system
gives way to a non-collinear spiral state after hole doping.
This spiral correlation is induced by a hole hopping term.
The competition between the AF correlation
and the spiral correlation leads to a modified hopping term
that has an SU(2) matrix form in spin space.
The components of the SU(2) matrix are determined by
the spin configuration along the hopping direction.\cite{Korenman1977}
In the continuum, one can see that this hopping term
scales as $1/\alpha$.
However, it turns out that just having this term is not
sufficient to stabilize a skyrmion.

In this paper, we fully account for quantum effects
in the formation of Zhang--Rice singlets
in the single-hole-doped CuO$_2$ plane to show that a skyrmion spin texture
is formed around the singlet.
The crucial point is that we need {\it both} the presence of 
a Zhang--Rice singlet and the hopping effect favoring a spiral state.

This paper is organized as follows.
In \S \ref{sec_model}, we introduce the model
that describes the single-hole-doped CuO$_2$ plane.
In \S \ref{sec_result}, we present the numerical
diagonalization result that suggests 
that a half-skyrmion spin texture is formed around
a Zhang--Rice singlet.
In \S \ref{sec_sc}, we present the self-consistent
skyrmion configuration.
In \S \ref{sec_discussion}, the physical consequences
of the skyrmion picture are discussed.
Section \ref{sec_conclusion} is devoted to the conclusion.

\section{Model}
\label{sec_model}
We consider a doped hole in the CuO$_2$ plane
in a skyrmion spin texture background.
In the single-hole-doped CuO$_2$ plane,
the system gains energy from Cu-O hybridization
that leads to an AF interaction 
between O and Cu holes.\cite{ZhangRice1988}
Zhang and Rice\cite{ZhangRice1988} considered the combinations of
the four O-hole states around a Cu ion, and showed
that the singlet state between a Cu hole and
a symmetric O-hole state with respect to the central Cu ion
has the largest binding energy.
By projecting out the other O-hole states, the $t$--$J$ model
was derived.\cite{ZhangRice1988}
Although we assume that the correlation of forming this singlet
state is the most important electronic correlation,
we do not restrict ourselves to the Hilbert space
where the singlet is fully formed.
The Hamiltonian is
\begin{eqnarray}
{\cal H} &=&
- t\sum_{\left\langle {\ell ,\ell '} \right\rangle, \sigma }
{\left( {p_{\ell' \sigma }^\dag  p_{\ell \sigma }
+ {\rm h.c.}
} \right)}
+ 2 J_K {\bf S}_0 \cdot 
\left( {\psi _0^\dag  {\boldsymbol \sigma} \psi _0 } \right) \nonumber \\
& &  + \frac{J_K}{2} \sum\limits_{j\left( { \ne 0} \right),\ell }
{{\bf{S}}_j  \cdot } \left( {p_\ell ^\dag  {\boldsymbol \sigma} p_\ell  } \right)
+ J \sum_{\langle i,j \rangle} {\bf S}_i \cdot {\bf S}_j,
\label{eq_H}
\end{eqnarray}
Here, the indices $\ell$ and $\ell'$ refer to O sites,
and $i$ and $j$ refer to Cu sites.
The hole hopping integral $t$ is restricted to nearest-neighbor sites,
and the components of ${\boldsymbol \sigma}=(\sigma_x, \sigma_y, \sigma_z)$
are the Pauli matrices.
The exchange interaction $J$ between Cu spins ${\bf S}_j$ 
is restricted to nearest-neighbor sites.
The operator $p_{\ell \sigma}^{\dagger}$ creates
an O ($2p_x$, $2p_y$) hole with spin $\sigma$ at site $\ell$.
In order to describe Zhang--Rice singlet formation,
the Cu spin at the center is described by the quantum spin
\be
{\bf S}_0 = \frac{1}{2}\left( {d_0^\dag  {\boldsymbol \sigma} d_0} \right),
\ee
with $d_0^\dag = (d_{0\uparrow}^\dag, d_{0\downarrow}^\dag)$.
The operator $d_{0\sigma}^\dag$ creates a Cu $3d_{x^2-y^2}$
hole with spin $\sigma$ at the central Cu site.
The second term of eq.~(\ref{eq_H}) describes the interaction
of forming a Zhang--Rice singlet.
The operator $\psi_0$ consists of four O-hole states around the
quantum Cu spin at the center and is given by
\be
\psi_{0 \sigma} = \frac{1}{2}\sum_{\ell \in \{0\}} p_{\ell\sigma}.
\ee
If we take a strong $J_K$ limit,
the Zhang--Rice singlet is fully formed and
eq.~(\ref{eq_H}) is reduced to the $t$-$J$ model.
However, here, we do not take this limit.

We assume a skyrmion configuration for ${\bf S}_j$
except at the origin ($j=0$).
By representing the coordinate vector of the Cu site $j$ by ${\bf R}_j$,
${\bf S}_j$ is given by
\be
{\bf S}_j = (-1)^j S {\bf n}_{sk}({\bf R}_j,\lambda),
\label{eq_Sj_sky}
\ee
where ${\bf n}_{sk}$ is defined by eq.~(\ref{eq_sky}).
For $S$, we set $S=1/2$.
(If we take into account the effect of quantum fluctuations,
$S$ is somewhat reduced.
However, we neglect this for simplicity.)
We use NL$\sigma$M in evaluating
the energy associated with the last term in eq.~(\ref{eq_H}).
The energy of a spin texture in NL$\sigma$M is given by 
eq.~(\ref{eq_Esky}).
The salient feature of this energy formula is that the energy 
is independent of the skyrmion size $\lambda$.
The resulting spin texture can be different from the originally assumed 
skyrmion configuration because of the presence of a quantum Cu spin 
at the origin.
However, the static Cu spin configuration is still characterized 
by the topological charge $Q$,
and the energy is given by eq.~(\ref{eq_Esky}) as long as 
we consider the lowest energy state in the homotopy sector with $Q$.
Since eq.~(\ref{eq_H}) is independent of the skyrmion size $\lambda$,
we may focus on the energy of the hole and 
the quantum spin ${\bf S}_0$ and
their interaction with other Cu spins.
Therefore, the Hamiltonian eq.~(\ref{eq_H}) is reduced to the following form:
\begin{eqnarray}
{\cal H} &=&
- t\sum_{\left\langle {\ell ,\ell '} \right\rangle, \sigma }
{\left( {p_{\ell' \sigma }^\dag  p_{\ell \sigma }
+ {\rm h.c.}
} \right)}
+ J_K \left( {d_0^\dag  {\boldsymbol \sigma} d_0} \right) \cdot 
\left( {\psi _0^\dag  {\boldsymbol \sigma} \psi _0 } \right) \nonumber \\
& &  + \frac{J_K}{2} \sum\limits_{j\left( { \ne 0} \right),\ell }
{{\bf{S}}_j  \cdot } \left( {p_\ell ^\dag  {\boldsymbol \sigma} p_\ell  } \right)
+ \frac{J}{2} \sum_{\langle 0,j \rangle} 
\left( {d_0^\dag  {\boldsymbol \sigma} d_0} \right) \cdot {\bf S}_j 
+ E_{NL{\sigma}M}.
\label{eq_H2}
\end{eqnarray}
In the second last term, we consider the interaction between ${\bf S}_0$ 
and its nearest neighbor Cu spins.
In the analysis below, we denote energies in units of $t$.

\section{Skyrmion Creation around Zhang--Rice Singlet}
\label{sec_result}
\begin{figure}[t]
   \begin{center}
     \includegraphics[width=0.9 \linewidth]{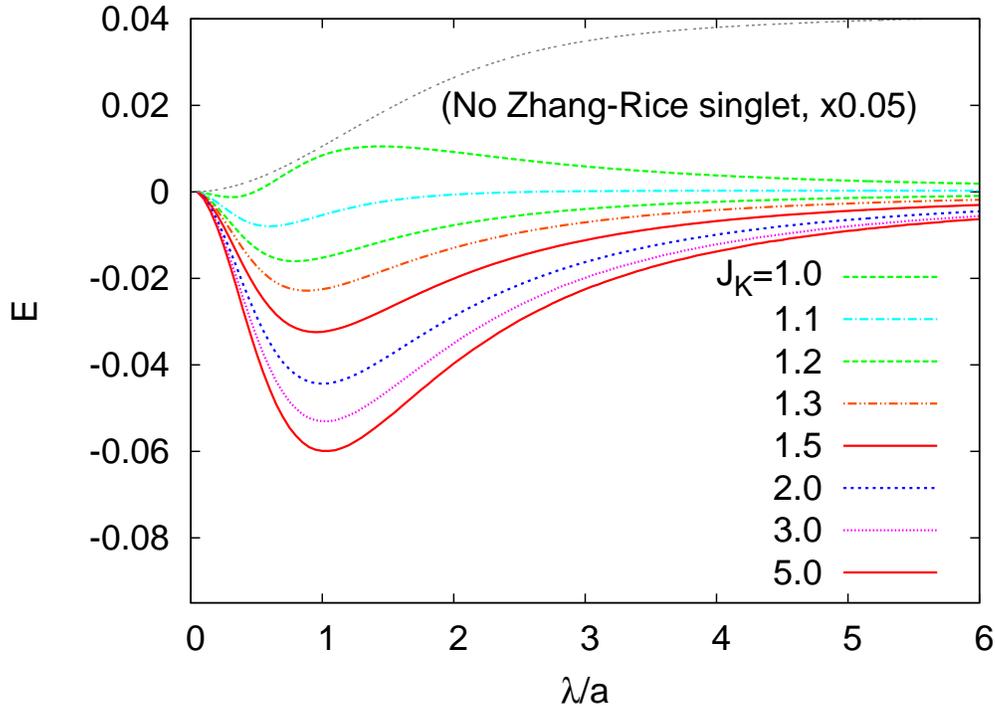}
   \end{center}
   \caption{ \label{fig_energy}
     (Color online)
     Skyrmion core size $\lambda$ dependences of the ground-state
     energy of the doped hole Hamiltonian eq.~(\ref{eq_H})
     for different values of $J_K$.
     The energy without a Zhang--Rice singlet for $J_K=2$
     is also shown.
     The energies are measured from their values at $\lambda=0$.
     In describing the CuO$_2$ plane,
     we employed a $21 \times 21$ square lattice for Cu sites
     and a $12 \times 12$ square lattice for O sites
     with open-boundary conditions.
    }
 \end{figure}

The Hamiltonian eq.~(\ref{eq_H2}) is exactly diagonalized
in the Hilbert space where each of the central Cu-hole
and the O-hole states is singly occupied.
For $J$, we assume $J/t=1/3$,
which is a standard value taken from the literature.\cite{LeeNagaosaWen2006}
The $\lambda$ dependence of the ground-state energy
is shown in Fig.~\ref{fig_energy}.
The ground-state energy
exhibits a clear energy minimum for $J_K>1$.
In particular, the minimum is located at $\lambda \simeq a$,
and this value is insensitive to an increase in $J_K$
for $J_K>1.5$.

Now, we estimate $J_K$.
From the second-order perturbation theory with respect to 
the Cu-O hybridization term,\cite{ZhangRice1988} 
$J_K$ is given by
\be
J_K = \frac{2 t_{dp}^2}{U_d - \Delta } 
+ \frac{2 t_{dp}^2}{U_p + \Delta },
\ee
where $t_{dp}$ is the wave-function overlap of Cu and O holes.
$U_p$ is the on-site Coulomb repulsion at an O site
and $U_d$ is the on-site Coulomb repulsion at a Cu site.
The parameter $\Delta$ is given by 
$\Delta  = {\varepsilon _p} - {\varepsilon _d}$,
with $\varepsilon_p$ and $\varepsilon_d$ being
the atomic energies of the O and Cu holes, respectively.
Using the parameter set taken from constrained
local density approximation calculations\cite{Hybertsen1989},
we find $J_K \simeq 2$.
Therefore, a skyrmion with a core size of $\lambda \simeq a$
is formed around the Zhang--Rice singlet.
Since the core state is the Zhang--Rice singlet state,
where there is no topological charge density,
this result suggests that a {\it half-skyrmion} with $|Q|=1/2$
is created around the Zhang--Rice singlet.
It is important to note that the presence of the Zhang--Rice singlet
is crucial to stabilization of the skyrmion.
Thus, if one replaces the quantum Cu spin with the classical spin,
the minimum disappears, as shown in Fig.~\ref{fig_energy}.

The location of the energy minimum at $\lambda \simeq a$
can be explained by a simple calculation.
Let us consider a cluster consisting of a single
Cu site at $(0,0)$ and four surrounding O sites
at $(\pm a/2,0)$, $(0,\pm a/2)$.
We assume that there are molecular fields at the O sites
created by other Cu spins at $(\pm a,0)$ and $(0,\pm a)$.
Then, the exact diagonalization of this cluster Hamiltonian
is carried out analytically.
(The details of the calculation is presented in Appendix \ref{app_mfe}.)
For the radial configuration of the molecular fields,
which is equivalent to the skyrmion configuration with $\lambda=a$,
we find that the energy is given by
\be
E_r =  - \frac{3J_K  + 2t}{2}
- \frac{1}{2}\sqrt {\left( {3J_K  + 2t} \right)^2  + S^2 J_K^2 }.
\label{eq_Er}
\ee
Meanwhile, for the AF configuration of the molecular fields,
we find 
\be
E_{AF} =  - J_K  - 2t
- \sqrt {\left( {2J - \frac{J_K}{2}} \right)^2 S^2 + 4J_K^2 }.
\label{eq_EAF}
\ee
Using the parameters for the CuO$_2$ plane,
we find $E_r < E_{AF}$ as shown in Fig.~\ref{fig_Ec}.
This can be simply understood as follows:
In order to maximize the energy gain by creating
the Zhang--Rice singlet state, the system prefers
to have the molecular fields cancel each other.
This is achieved for the radial configuration.
By contrast, the molecular field effect is maximized
for the AF configuration.

The instability of the homotopy sector with $Q=0$, which 
corresponds to the AF state,
is understood in the following way as well.
In order to clarify the physical picture,
we rewrite the NL$\sigma$M by introducing
complex fields $z_\sigma  \left( {\bf r},t \right)$
through 
\be
{\bf n}({\bf r},t)=\sum_{\sigma,\sigma'=\uparrow,\downarrow}
z_{\sigma}^{*} ({\bf r},t) 
{\boldsymbol \sigma}_{\sigma,\sigma'} z_{\sigma'} ({\bf r},t),
\ee
and we obtain the CP$^1$ model\cite{Rajaraman}
\bea
S_{CP^1} &=& \frac{4}{g} 
\int dx \int dy \int d (c_{\rm sw}t) 
\sum_{\sigma = \uparrow, \downarrow} 
\left[
  \left| 
  \left( 
  \frac{1}{c_{\rm sw}} \partial _{t} - i\alpha _0
  \right) 
  z_\sigma  \left( {\bf r},t \right) \right|^2 
  \right. \nonumber \\
  & & \left.  
  - \left| \left( {\partial _{x}   + i\alpha _x } \right)
  z_\sigma  \left( {\bf r},t \right) \right|^2
  - \left| \left( {\partial _{y}   + i\alpha _y  } \right)
  z_\sigma  \left( {\bf r},t \right) \right|^2
  \right],
\eea
with
\be
\alpha _0   = i \sum_{\sigma = \uparrow, \downarrow}
z_\sigma ^* \left( {\bf r},t \right) 
\frac{\partial}{\partial {\left( c_{\rm sw} t \right) }}
z_\sigma \left( {\bf r},t \right),
\ee
\be
{\boldsymbol \alpha } = \frac{i}{2} \sum_{\sigma = \uparrow, \downarrow}
\left\{
z_\sigma ^* \left( {\bf r},t \right) 
\nabla z_\sigma \left( {\bf r},t \right)
-
\left[ \nabla z_\sigma ^* \left( {\bf r},t \right) \right]
z_\sigma \left( {\bf r},t \right)
\right\}.
\ee
The equation of motion for the steady state is given by
\be
\left( \nabla  + i \boldsymbol{\alpha} \right)^2 z_\sigma = 0.
\label{eq_za}
\ee
In the presence of the Zhang--Rice singlet, the boundary condition
at the origin is 
\be
z_{\sigma} (0) = 0.
\label{eq_z_bc}
\ee
In the AF configuration, there is no gauge flux, 
$\nabla \times {\boldsymbol \alpha} = 0$.
Therefore, the equation of motion is reduced 
to
\be
\nabla^2 z_\sigma = 0.
\ee
Using the polar coordinate $r$ and $\phi$ with 
$x=r\cos \phi$ and $y=r\sin \phi$,
let $z_\sigma = f_\sigma (r) \exp(i \ell \phi)$.
Here, $\ell$ is an integer.
The equation for $f_\sigma (r)$ is given by
\be
\frac{d^2 f_\sigma}{dr^2} + \frac{1}{r}\frac{df_\sigma}{dr}
-\frac{\ell^2}{r^2} f_\sigma = 0.
\ee
To be consistent with the assumption that there is no gauge flux,
we must take $\ell=0$.
The equation is readily solved and we find
\be
f_{\sigma} \propto \ln r.
\ee
Clearly, this solution is incompatible with
the boundary condition eq.~(\ref{eq_z_bc}).

In the case that there is a nonvanishing gauge flux,
it is difficult to solve the equation of motion because
it is a nonlinear equation with respect to $z_\sigma ({\bf r},t)$.
However, it is possible to find the behavior of the solution
near $r=0$.
If we assume $z_\sigma = f_\sigma (r) \exp(i \ell \phi)$
with $\ell \neq 0$, we find
\be
{\boldsymbol \alpha}  =  - \frac{\ell}{r} \sum_{\sigma}  
{f_\sigma ^2} {{\bf e}_\phi },
\ee
where ${\bf e}_\phi = (-\sin \phi, \cos \phi)$.
From eq.~(\ref{eq_za}), after some algebra, we obtain
\be
\frac{{{d^2}{f_\sigma }}}{{d{r^2}}} 
+ \frac{1}{r}\frac{{d{f_\sigma }}}{{dr}} 
- \frac{{{\ell ^2}}}{{{r^2}}}
\sum_{\sigma '} {{{\left( {1 - f_{\sigma '}^2} \right)}^2}} {f_\sigma } = 0.
\ee
For $r \rightarrow 0$, the solution of this equation 
behaves as
\be
f_\sigma \rightarrow C_\sigma r^{|\ell|},
\ee
with $C_\sigma$ as a constant.
Note that this form is consistent with 
the boundary condition eq.~(\ref{eq_z_bc}).
Since $\ell \neq 0$, there is a gauge flux.
The situation is quite similar to the problem of the point
singularity in a two-dimensional superfluid.\cite{FetterWalecka}
The vortex in $z_{\sigma}({\bf r})$ turns out to be a skyrmion.
In fact the topological charge density eq.~(\ref{eq_tcd})
is expressed as
\be
\rho_{tc}({\bf r})= -\frac{1}{2\pi} 
\left( \partial_x \alpha_y - \partial_y \alpha_x \right),
\ee
in terms of the gauge flux.\cite{Rajaraman}
Therefore, a nonvanishing gauge flux is a skyrmion
characterized by the topological charge eq.~(\ref{eq_Q}).

\section{Self-Consistent Skyrmion Configuration}
\label{sec_sc}
In the previous section, we have presented the numerical result
that suggests that a half-skyrmion spin texture is formed around
the Zhang--Rice singlet.
In this calculation, Cu spins except that at the origin
are assumed to be a skyrmion spin texture given by
eq.~(\ref{eq_Sj_sky}).
In this section, we attempt to obtain the self-consistent 
spin configuration by fixing the directions
of the four Cu spins surrounding the Zhang--Rice singlet.

\begin{figure}[t]
   \begin{center}
     \includegraphics[width=0.95 \linewidth]{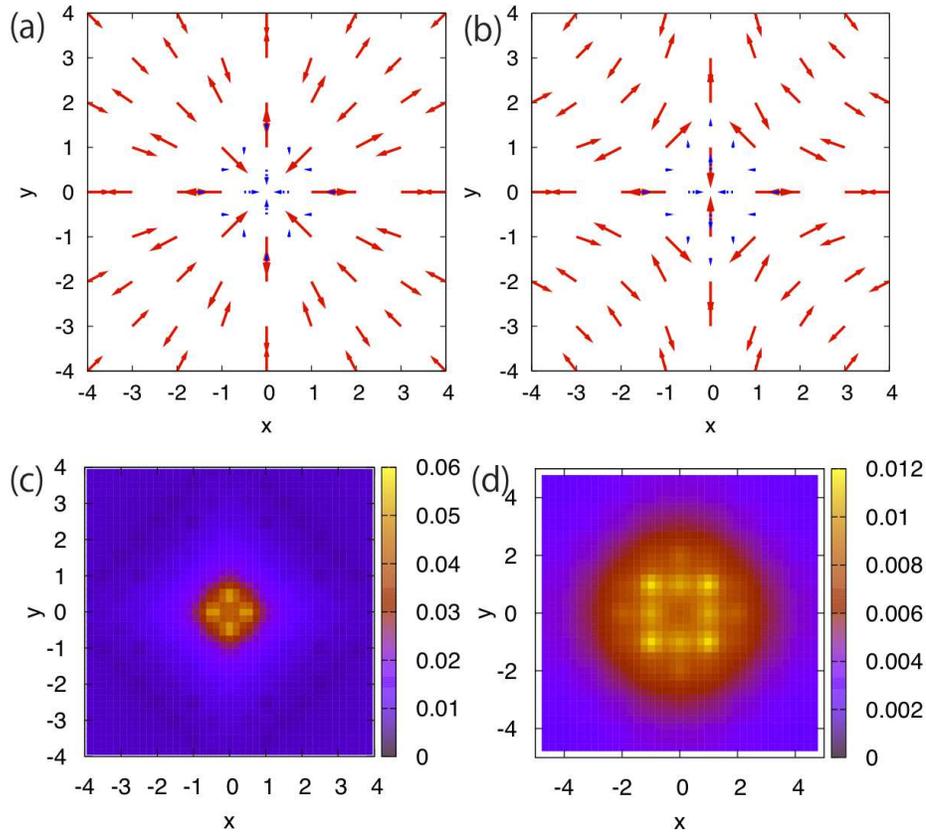}
   \end{center}
   \caption{ \label{fig_sc}
     (Color online)
     Self-consistent Cu spin configuration
     in the case of a skyrmion (a) and an anti-skyrmion (b).
     The dashed arrows represent the doped hole spin, which is scaled
     by a factor of 10.
     (c) Doped hole density and
     (d) topological charge density
     for the skyrmion (a).
    }
 \end{figure}

The self-consistent Cu spin configuration is obtained
by solving the mean field equations.
At each Cu site, there are molecular fields
created by the nearest-neighbor Cu sites and O-hole.
In the self-consistent configuration, the Cu spin at each site
is antiparallel to this molecular field.
To describe the CuO$_2$ plane,
we employ a $15 \times 15$ square lattice for Cu sites and
a $10\times 10$ square lattice for O sites
under the open-boundary condition.
As explained below, the doped hole is strongly localized around
the Zhang--Rice singlet.
Therefore, a large system is not necessary in describing
doped hole hopping.
According to the energy calculation in Fig.~\ref{fig_energy}
and the cluster analysis above, we fix the four Cu spins
around the center.
As shown in the self-consistent spin configurations in Fig.~\ref{fig_sc},
the difference between the skyrmion and the anti-skyrmion
is only that the $x$-components of the spins are reversed.
Furthermore, as shown in Fig.~\ref{fig_sc}(c), the doped hole
is strongly localized around the quantum Cu spin.
In the absence of the Zhang--Rice singlet,
the doped hole is not localized but extended.
Therefore, the formation of a skyrmion is not sufficient to
localize the doped hole.

The physically important quantity is the topological charge
density shown in Fig.~\ref{fig_sc}(d).
Note that rotation of all spins about the $z$-axis through the same angle
does not change the topological character.
However, the topological charge changes sign for the anti-skyrmion case.
It is important to note that the topological charge density
is considerably extended when compared with the doped hole wave function.
Since the topological charge distribution produces
a magnetic-field-like effect through the Berry phase,\cite{Berry1984}
the non-vanishing topological charge density carried by
the doped hole produces intriguing effects,
as will be discussed below.
The sum of the topological charge is $0.320$,
whereas the expected value is one-half.
This discrepancy is probably associated with the approximation
in describing Cu spins by classical spins.
In fact, the calculation above overestimates the AF correlation
that reduces the total topological charge.
Describing Cu spins using more suitable fields
is a future problem.

The existence of a stable skyrmion spin texture in a Li-doped system
was suggested by Haas {\it et al}.\cite{Haas1996}
in the $t-J$ model with one hole on a 13-site cluster.
In the cluster, the central Cu site is replaced by Li.
The exact diagonalization of the cluster system
under the boundary condition imposed by
the skyrmion configuration shows that
the total energy has a minimum at $\lambda \simeq 2a$.
In their calculation, the doped hole cannot reside
at the center site and is bounded
by the attractive Coulomb potential of Li.
Although the system is not equivalent to our system,
the mechanism of stabilizing a skyrmion is
quite similar.
In the model in ref.~\citenum{Haas1996},
the doped hole is localized owing to the attractive
Coulomb potential.
In our system, however, the doped hole is localized by the formation of
the Zhang--Rice singlet.
The stability of a skyrmion is also demonstrated 
in a 16-site cluster\cite{Gooding91}
with hole motion being restricted to four sites around the center.

\section{Discussion}
\label{sec_discussion}
Now, we explore the physical consequences of the skyrmion spin texture created by a doped hole.
The most direct consequence is that the magnetic properties
are straightforwardly connected with the hole doping concentration $x$.
As revealed by neutron scattering measurements,\cite{Keimer1992}
the inverse of the AF correlation length $\xi_{AF}$ is well described by
the simple formula 
\be
\kappa (x,T)=\kappa (x,0) + \kappa (0,T),
\ee
where $T$ is the temperature.
The inverse of the temperature-independent term $\kappa (x,0)$
is given by
\be
\frac{1}{\kappa (x,0)} = \frac{a}{\sqrt{x}}.
\ee
The right-hand side of this equation is 
the average separation between the doped holes.
Meanwhile, $\kappa (0,T)$ is described 
by the formula for NL$\sigma$M.\cite{Hasenfratz1991}
The latter is associated with the underlying AF correlation.
When each doped hole creates a skyrmion,
local spin disorder is created by the skyrmion.
Therefore, the average hole separation leads to
a temperature-independent component of $\xi_{AF}$.
Another important observation is that the skyrmion character
is lost at approximately around $x \sim 0.3$, 
where the average hole separation
is $\sim 2a$.
In the vicinity of this critical doping level, 
each doped hole is unable to
create a skyrmion with $\lambda = a$ and so all skyrmion features 
are lost for $x>0.3$.

As we have seen above, the doped hole is localized
by the formation of the Zhang--Rice singlet.
Therefore, Nagaoka ferromagnetism\cite{Nagaoka1966} does not occur
and the system is insulating in a slightly doped regime.
The energy band of doped holes forms when
the overlap of adjacent doped hole wave functions
is not negligible.
The bandwidth $W$ is proportional to the wave function overlap.

As mentioned above, the presence of a skyrmion leads to a Berry phase effect.
The skyrmion produces a magnetic-field-like effect for
doped holes. When a doped hole enters a region of
skyrmion spin texture created by another doped hole, then the Lorentz force
effectively acts between the two holes.
This interaction leads to a chiral pairing state.\cite{Morinari2006}
Since there are skyrmions and anti-skyrmions,
two chiral pairing states coexist with opposite chiralities.
The analysis of the gap equation suggests that
this type of interaction leads to $d_{x^2-y^2}$-wave
superconductivity.\cite{Morinari2006,MorinariMFS2010}

In addition, skyrmion dynamics has a nontrivial contribution 
to hole dynamics.
The skyrmion itself has its own dynamics, which is not
identical to bare hole dynamics.\cite{Morinari2005}
The doped hole moves with a constraint of keeping the skyrmion
as long as $W < J$.
The underlying Fermi surface, which appears when the skyrmion
is suppressed, is partially gapped because the skyrmion
energy does not necessarily vanish along it, which results
in a truncated Fermi surface.\cite{MorinariMFS2010}
For this pseudogap phenomenon, the characteristic
temperature is the binding energy of the skyrmion,
which is $\sim 0.04t$ from Fig.~\ref{fig_energy}.
The order of magnitude for this value is the same as that of the pseudogap
temperature evaluated in materials with high $T_c$ cuprates.\cite{Timusk1999}
However, for elaborate investigations of the pseudogap,
a better description is necessary.
In this regard,
it is worth pointing out that the Hamiltonian
describing the dynamics of the composite object comprising
the doped hole and skyrmion is similar
to that for quasiparticles
in the $d$-density wave state.\cite{Chakravarty2001}
However, the physical picture is quite different.

Here, we comment on electron-doped high-$T_c$ cuprates.
The superconducting state of electron-doped cuprates has 
a $d$-wave pairing symmetry,\cite{TsueiKirtley2000} which is the same
as that in the hole-doped case.
From this observation one might expect that the mechanisms
of superconductivities are the same for electron-doped
and hole-doped cuprates.
However, the superconducting transition temperature 
of electron-doped cuprates is quite low compared with
that of hole-doped cuprates with high transition temperatures.
Furthermore, there are qualitative differences between
electron-doped and hole-doped cuprates.
Among others, the relationships between superconductivity 
and antiferromagnetism are strikingly different.
In hole-doped cuprates, only $2-3\%$ hole doping is sufficient
to kill AF order, while we need about $20\%$ electron doping
to suppress AF order in electron-doped cuprates.
In the phase diagram in the plane of doping concentration
and temperature, the superconducting phase is clearly separated 
from the AF phase in hole-doped cuprates, while
the superconducting phase and the AF phase sit side by side
in the electron-doped cuprates.
\cite{ShenRMP}
Therefore, it is quite natural to surmise a different mechanism 
for superconductivity works in electron-doped cuprates.
If the mechanism is different, it is quite hard to describe
such a difference in the single-band Hubbard model.
Meanwhile, the difference is simply associated with
the presence or absence of skyrmions in the skyrmion picture.
In electron-doped cuprates, Zhang-Rice singlets are irrelevant
because electron carriers are introduced at Cu sites and there is 
no singlet correlation between electron
carriers and Cu $d$-electrons.
Therefore, a skyrmion is not formed in electron-doped cuprates.

Finally, we propose a ``smoking-gun'' experiment
to establish the presence of skyrmions in high-$T_c$ cuprates.
In the doping range of $0.02 < x < 0.055$,
incommensurate inelastic magnetic peaks are observed
in neutron-scattering experiments.\cite{Cheong1991,Yamada1998}
In the skyrmion picture, these peaks are associated
with the AF skyrmion lattice.\cite{MorinariMFS2010}
Recently, real-space observations of a two-dimensional
skyrmion lattice have been reported in Fe$_{0.5}$Co$_{0.5}$Si.\cite{Yu10}
This experimental technique can be applied to finding
the skyrmion lattice in high-$T_c$ cuprates as well.

\section{Conclusions}
\label{sec_conclusion}
To conclude, we have shown that
a skyrmion spin texture is created around a Zhang--Rice singlet
in the single-hole-doped CuO$_2$ plane.
The evolution of the AF correlation upon hole doping
is naturally described by the skyrmion picture.
A doped hole with a skyrmion creates the Berry phase effect,
which can lead to $d$-wave superconductivity.
The skyrmion picture also seems to be consistent
with some other aspects of high-$T_c$ cuprates.

\begin{acknowledgments}
I would like to thank S. Brazovski, N. Nakai, and L. Cai
for discussions and T. Tohyama for comments.
This work was supported by
a Grant-in-Aid for the Global COE Program
``The Next Generation of Physics, Spun from Universality and Emergence''
from the Ministry of Education, Culture, Sports, Science
and Technology (MEXT) of Japan.
\end{acknowledgments}

\appendix
\section{Derivation of Eqs.~(\ref{eq_Er}) and (\ref{eq_EAF})}
\label{app_mfe}
\begin{figure}[t]
   \begin{center}
     \includegraphics[width=0.9 \linewidth]{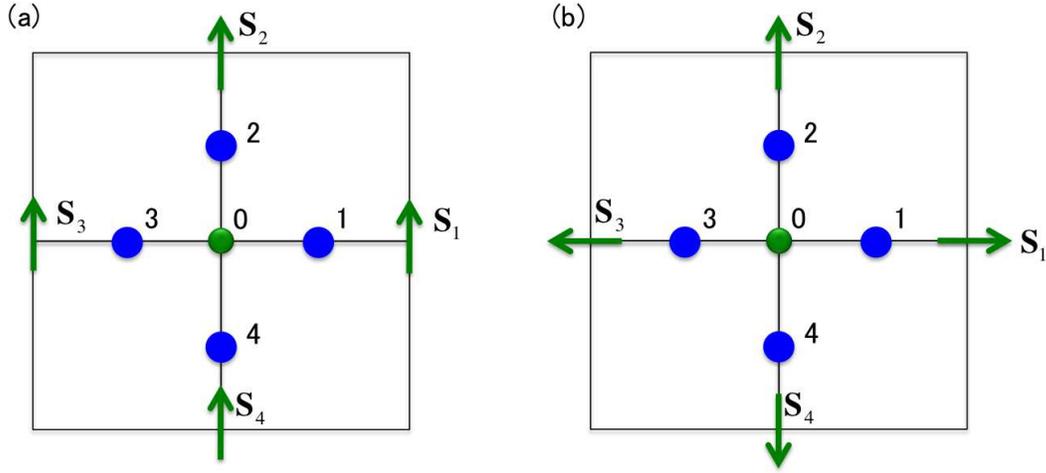}
   \end{center}
   \caption{ \label{fig_mfe}
     (Color online)
     Molecular fields ${\bf S}_j$ ($j=1,2,3,4$)
     at Cu sites for the (a) AF configuration 
     and (b) radial configuration.
     The circle at the center denotes the quantum Cu spin.
     The O sites are labeled as $1,2,3$, and $4$.
   }
 \end{figure}

In this Appendix, we present the derivation of eqs.~(\ref{eq_Er})
and (\ref{eq_EAF}).
We consider a cluster that consists of one quantum Cu spin
sitting at the center, four O sites,
and four classical Cu spins ${\bf S}_j$,
as shown in Fig.~\ref{fig_mfe}.
For convenience, we introduce the following field:
\be
   {\gamma _{q \sigma }}
   = \frac{1}{2}\sum_{\ell  = 1,\,2,\,3,\,4} 
   {\exp \left( {i q \left( {\ell  - 1} \right)} \right){p_{\ell \sigma }}},
\ee
where $q=0,\pi/2,\pi,3\pi/2$.
In terms of these fields, 
the Hamiltonian of the cluster is given by
\bea
    {\cal H}
    &=&  2t \sum_{\sigma} \left( 
    -\gamma _{0\sigma}^\dag {\gamma _{0\sigma}} 
    + \gamma _{\pi\sigma} ^\dag {\gamma _{\pi \sigma} } \right)
    + {J_K}\left( {d_0^\dag {\boldsymbol \sigma} {d_0}} \right) 
    \cdot \left( {\gamma _0^\dag {\boldsymbol \sigma} {\gamma _0}} \right) 
    \nonumber \\
    & & + \frac{{{J_K}}}{2} \sum_{j,\,p,\,q} 
              {{\left( {{M_j}} \right)}_{pq}}
              {{{\bf{S}}_j} \cdot 
                \left( {\gamma _p^\dag {\boldsymbol \sigma} {\gamma _q}} 
                \right)}  
              + \frac{J}{2}\sum_j {{{\bf{S}}_j} 
                \cdot \left( {d_0^\dag {\boldsymbol \sigma} {d_0}} \right)},
\label{eq_Hc}              
\eea
where $(M_j)_{pq} = (U^{\dagger})_{pj} U_{jq}$ with
\be
U = \frac{1}{2}\left( {\begin{array}{cccc}
1 & 1 & 1 & 1\\
i & -i & 1 & -1\\
-1 & -1 & 1 & 1\\
-i & i & 1 & -1
\end{array}} \right).
\ee

For the AF configuration of ${\bf S}_j$ (Fig.~\ref{fig_mfe}(a)),
we take ${\bf S}_j = S {\bf e}_z$, where ${\bf e}_z$
is the unit vector in the $z$-direction.
The third term in eq.~(\ref{eq_Hc}) has the following simple form:
\be
\frac{{S{J_K}}}{2}
\sum_p {\gamma _p^\dag {\sigma _z}{\gamma _p}}.
\ee
In the Hilbert space, where each of the central Cu states and O states
is singly occupied, the Hamiltonian is diagonalized analytically,
and we find that the ground-state energy is given by eq.~(\ref{eq_EAF}).

For the radial configuration of ${\bf S}_j$ [Fig.~\ref{fig_mfe}(b)],
we take ${\bf S}_1 = S {\bf e}_x$,
${\bf S}_2 = S {\bf e}_y$,
${\bf S}_3 = - S {\bf e}_x$,
and ${\bf S}_4 = - S {\bf e}_y$.
Here, ${\bf e}_x$ and ${\bf e}_y$
are the unit vectors in the $x$- and $y$-directions,
respectively.
In this case, the third term in eq.~(\ref{eq_Hc}) has the form
\be
\frac{{S{J_K}}}{2}\left( {\gamma _{\pi /2}^\dag {\sigma _ - }{\gamma _0} 
+ \gamma _{ - \pi /2}^\dag {\sigma _ + }{\gamma _0} 
+ \gamma _0^\dag {\sigma _ + }{\gamma _{\pi /2}} 
+ \gamma _0^\dag {\sigma _ - }{\gamma _{ - \pi /2}}} \right),
\ee
where $\sigma_{\pm} = (\sigma_x \pm i \sigma_y)/2$.
Noting that this term does not contain $\gamma_{\pi \sigma}$
or $\gamma_{\pi \sigma}^{\dagger}$,
we find that the Hilbert space is divided into subspaces.
Explicitly considering spin states of the Cu spin state
and the O hole state, the problem is reduced to diagonalize 
the following $4 \times 4$ matrix:
\be
\left( 
{\begin{array}{cccc}
{ - 2t - {J_K}}&{2{J_K}}&{\frac{{S{J_K}}}{2}}&0\\
{2{J_K}}&{ - 2t - {J_K}}&0&{\frac{{S{J_K}}}{2}}\\
{\frac{{S{J_K}}}{2}}&0&0&0\\
0&{\frac{{S{J_K}}}{2}}&0&0
\end{array}} \right).
\ee
By taking the lowest eigenenergy of this reduced Hamiltonian,
we obtain eq.~(\ref{eq_Er}).

\begin{figure}[t]
   \begin{center}
     \includegraphics[width=0.9 \linewidth]{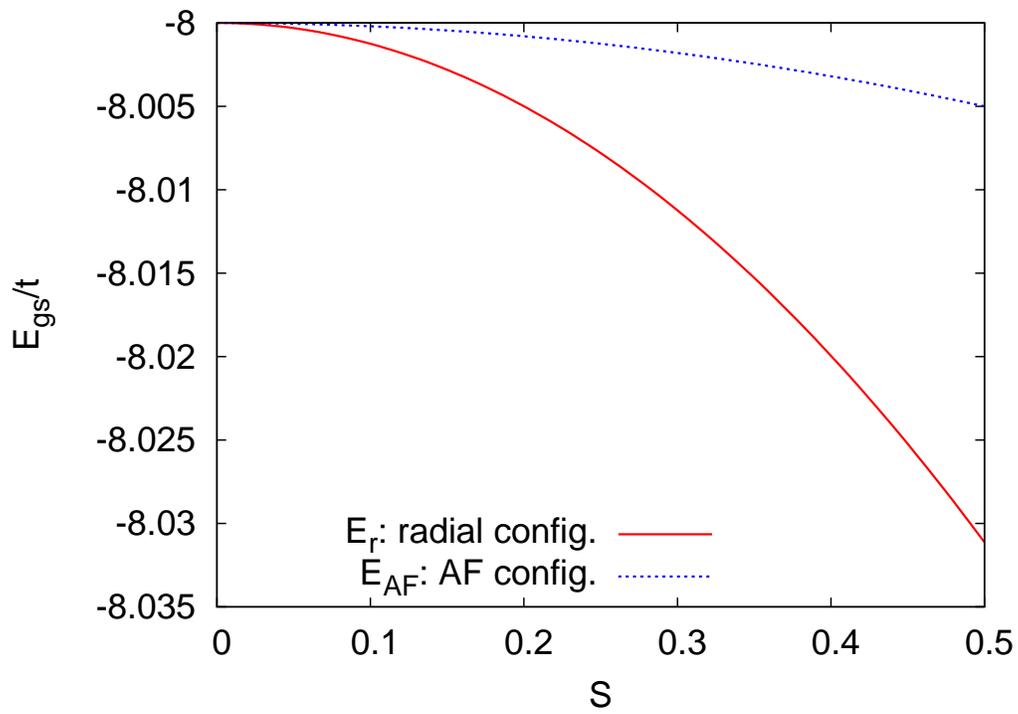}
   \end{center}
   \caption{ \label{fig_Ec}
     (Color online)
     $E_r$ and $E_{AF}$ as functions
     of the spin $S$ for $J_K/t=2$.
   }
 \end{figure}

The energies of eqs.~(\ref{eq_Er}) and (\ref{eq_EAF})
are shown in Fig.~\ref{fig_Ec} as functions of $S$.
Here, we assume $J_K/t=2$ and $J/t=1/3$.
In the entire range of $S$,
the energy of eq.~(\ref{eq_Er}) is lower than 
the energy of eq.~(\ref{eq_EAF}).
We find that $E_r < E_{AF}$ for $J_K/t > 0.66$.
For $J_K/t < 0.66$, $E_r > E_{AF}$ in the entire range of $S$.


\end{document}